\documentstyle[prb,aps,epsf,amstex,a4]{revtex}
\begin{document}
\draft
\title{The Randomly Driven Ising Ferromagnet\\ Part I:\\General Formalism and Mean Field Theory}
\author{Johannes Hausmann$^{\dagger}$ and P\'al Ruj\'an$^{\dagger,\$}$}
\address{Fachbereich 8 Physik$^{\dagger}$ and ICBM$^{\$}$, Postfach 2503,
         Carl von Ossietzky Universit\"at,
         D-26111 Oldenburg, Germany}
 
\date{\today}
\maketitle
\begin{abstract}
  \noindent
We consider the behavior of an Ising ferromagnet obeying the
Glauber dynamics under the influence of a fast switching, random
external field. After introducing a general formalism for
describing such systems, we consider here the mean-field theory.
A novel type of first order phase transition related to 
spontaneous symmetry breaking and dynamic freezing is found. 
The non-equilibrium stationary state has a complex structure,
which changes as a function of parameters from a singular-continuous
distribution with Euclidean or fractal support to an absolutely
continuous one. These transitions are reflected in both finite size effects
and sample-to-sample fluctuations.
\end{abstract}
\pacs{\small PACS numbers: 05.50+g 05.70.Jk 64.60Cn 68.35.Rh 75.10.H 82.20.M}

\section{Introduction}

The last decade has seen many advancements in the theory of
dynamic systems, both from a mathematically rigorous, as from a physically
oriented heuristic point of view. Most of these results have
been obtained for systems with few degrees of freedom, attempts
at handling nonequilibrium stationary states of
systems with macroscopically many degrees
of freedom  have been made only recently \cite{SS,GC1,GC2}. 
In this paper we 
propose a general theoretical framework for 
{\it strongly, randomly
driven} statistical physical systems and apply it to 
the Ising model in a random, dichotomic driving external field. 

The resulting dynamics has many qualitative similarities with 
earlier work on
one dimensional Ising chains in a binary random field \cite{RFIM1}.
However, while the `dynamics' defined in \cite{RFIM2} is a one dimensional
map generated by the iteration of non-commuting $2\times 2$ transfer
matrices, in the present case the map `lives' in $2^N$ dimensions.
Nevertheless, the main mechanism leading to
chaotic behavior and strange attractors 
is in both cases related to the competition between 
two (or more) fixed points (or limit cycles).
In this respect, the randomly driven Ising model (RDIM) has many
intriguing aspects which -- somewhat unexpectedly -- can be
handled both analytically and numerically with methods developed earlier for
the random field Ising model \cite{RFIM2,evang,peter1,peter2,behn1,behn2,bene}.

The Ising  ferromagnet in a time-dependent sinusoidally oscillating field has received
recently a lot of attention, from both a theoretical and experimental
point of view. On the theoretical side, Rao, Krishnamurthy, and Pandit 
\cite{hyst-theo1} have
presented a large $N$-expansion of the cubic $O(N)$ model in three
dimensions and calculated the critical exponents related to the area of
the hysteresis loop. The underlying dynamic phase transition has been
then studied within both mean-field \cite{hyst-theo2} and Monte Carlo
simulations \cite{hyst-mc1,hyst-mc2,th-mc3,hyst-mc3}. The theory presented
in this paper is a generalization of these ideas for the case when
the external field is subject to a chaotic dynamics and/or is a random variable.

Besides the theoretical interest in describing such systems, 
we believe that
many of our predictions can be tested with recently developed experimental
techniques.
Dynamic magnetization measurements have been recently performed
in ultrathin Au(111)/Cu(0001)/Au(111) sandwiches 
or epitaxial Co/Au(111) films \cite{exp1,exp2,exp3}. Similarly,
hysteresis measurements on the ultrathin film
Co/Au(001) \cite{exp4} indicate that
below $T_c$ these systems undergo a dynamic phase transition
belonging to the Ising-universality class. 
More relevant to our theory, the time evolution 
of magnetization clusters can be optically recorded. The typical
relaxation times range from minutes to a few seconds with increasing
field amplitudes \cite{exp3}. 
This relatively slow relaxation rate allows for a simple 
experimental realization of the 
randomly driven external field.

Ultrathin films are potential candidates
for magneto-optical storage devices and our approach might be relevant 
especially in this respect. At well chosen control parameters the 
stationary magnetization distribution of the RDIM 
displays several well separated peaks. Thus, when driven appropriately,
such materials can store locally more than the two 
values typical for an equilibrium ferromagnetic system.

This first paper is organized as following: the basic assumptions and the
general theoretical formalism are introduced in Section~2. The mean-field
theory is presented in full detail in Section~3.
In this approximation the paramagnetic--ferromagnetic stationary phase
transition becomes first order.  The phase boundary
is obtained analytically. The average magnetization is non-analytic (jumps) at 
small driving field values and therefore the usual mean-field approach fails.
Nevertheless, the phase transition is related to a spontaneous symmetry
breaking and a pitchfork bifurcation of the magnetization distribution. 
On the other hand, the analytical nature of the stationary magnetization
distribution changes from singular continuous with euclidean or fractal support
to absolutely continuous along analytically computed boundaries.
Such changes are directly connected to finite size effects of the free energy
and the multifractal spectrum of the magnetization measure.
Close to the para-ferro phase boundary
the small magnetization region close to $m\sim 0$ becomes a repellor, 
inducing ultracritical slowing down related to type-I intermittency.
A short summary of the main mean-field features is presented in Section~4.

A subsequent paper considers the randomly driven Ising model in one and two dimensions.
Although the one-dimensional case cannot be fully solved,
many interesting exact results can be derived. We find
a line of second order phase transitions
at $T=0$ between a disordered (driven paramagnetic) and
an ordered (ferromagnetic) phase. 
Along this line the stationary critical exponents 
change continuously as a function of the driving field strength. 
However, the dynamic critical exponent
remains unchanged, $z=2$. As a function of temperature and of the
driving field strength, the nonequilibrium stationary state might
display multifractal or `fat' multifractal character. Hence, the
generalized free energy is characterized by anomalous
fluctuations related to the existence of a multifractal spectrum.
We also performed  Monte Carlo simulations on a square lattice.
Many features of the mean-field dynamics are shown
to survive the strong fluctuations characteristic to two dimensional systems.
However, in contrast to the mean-field approach, the two-dimensional model
displays also an interesting spatial structure related to droplet dynamics.
Comparisons between mean-field theory and two-dimensional results are
systematically presented, including some
preliminary results for hysteresis.


\section{General Formalism}
\subsection{The Master Equation}

Consider a spin system $\vec \mu =(s_1,s_2,\dots,s_N)$ in contact with a thermal
bath (phonons). Let us denote by $\tau_{spin-flip}$ the characteristic time of 
the spin-phonon interactions, and by $\tau_{sys}$ 
the {\it slowest} relaxation mode of the spin system.
For $\tau_{spin-flip} \ll \tau_{sys}$ the system is in {\it local} thermal
equilibrium.

Our basic assumption is that the
time evolution of the spin system can be described 
by a time-dependent joint probability distribution, $P(\{s_i\};t)$.
From a dynamic point of view, one can consider this distribution as
an average over all microstate initial conditions satisfying
given macroscopic constraints. The rigorous definition of this
basic assumption is related to the existence of Markov-partitions in
(hyperbolic) dynamic systems \cite{GC1,GC2} and goes beyond
the scope of this article.

The distribution $| P(t) \rangle$ obeys the Master Equation:
\begin{equation}
       | \dot P(t) \rangle  = - \hat {\cal L}_{B(t)}|  P(t) \rangle,
        \label{mastereq}
\end{equation}
where the operator $\hat {\cal L}_{B(t)}$ describes the outflow (inflow) probability
from (into) state $\{s_i\}$ into (from) other states, as prescribed by the 
dynamic rules.
The dependence on the field is made explicit because in
general $\left[ \hat {\cal L}_{B}, \hat {\cal L}_{B'}\right] \neq 0$ for $B\neq B'$. 
Assuming a constant field $B=B_0$ and
ordering the eigenvalues of $\hat {\cal L}_{B_0}$ as
$\lambda_0=0 \le |\lambda_1| \le \dots \le |\lambda_{2^N-1}|$, 
one has $\tau_{sys} = |\lambda_1|^{-1}$. 
The ket vector $|  P(t) \rangle$ 
can be expressed in the spin-configuration basis as $P(\{s_i\};t)$.
Alternately, it can be parameterized as
\begin{eqnarray}
        \label{Pparam}
         P(\{s_i\};t)  & = & {1\over 2^N} [ 1 + \sum_{i=1}^{N} m_i s_i + 
        \sum_{i \neq j} c_{i,j} s_i s_j + \dots ]  
         =  {1\over 2^N} \left( \sum_{\alpha=1}^{2^N} \pi_{\alpha} \prod_{j \in \alpha} s_j + 1\right),
\end{eqnarray}
where the $\pi_{\alpha}(t)$ are the average values of all possible $2^N$ products of spins
\begin{equation}
\pi_{\alpha}(t) = \langle \prod_{j\in \alpha} s_j \rangle = 
\sum_{\vec s} P(\{s_i\};t) \prod_{j\in \alpha} s_j
\label{pivec}
\end{equation}

Hence, $|P(t)\rangle$ can be expressed as a $2^N$-dimensional normalized vector in the
space of all possible spin configurations, or in the space of spin-products
as $\vec \pi (t) = (\langle s_1 \rangle,\dots, \langle s_N \rangle,
\langle s_1 s_2 \rangle, \dots)$. Note than when expressing the kinetic Ising model Liouville
operator in terms of Pauli matrices\cite{siggia} this orthogonal transformation
corresponds to the exchange $\sigma^z \leftrightarrow \sigma^x$.

In general, the Liouville operator $\hat {\cal L}$ is
not symmetric but can be expanded in a biorthogonal basis formed by its right, $|r_n \rangle$,
and left, $\langle l_n |$, eigenvectors ($\langle l_n |r_m \rangle \sim \delta_{m,n}$):
\begin{equation}
        \hat {\cal L} = \sum_{n=0}^{2^N -1} |r_n\rangle \lambda_n \langle l_n |
        \label{specdec}
\end{equation}
It is worth noting that since $e^{-\hat {\cal L}}$ is a stochastic operator, we have
$ \langle l_0 | \hat {\cal L} = 0 \langle l_0|$, where $ \langle l_0 | = (1,1,\dots, 1)$ in
the spin-configuration basis. 
The scalar product $\langle l_0 | r_0 \rangle = Z$ delivers the equilibrium (stationary)
canonical partition function.

The time-dependent field $B(t)$ is usually a deterministic one-dimensional
map, {\it e.g.} a harmonically oscillating field.
However, if the deterministic map is chaotic, the field 
becomes a random variable. In what follows we assume
that the external field $B$
is a random variable sampled identically and independently from the symmetric
distribution $\rho(B) = \rho(-B)$.

Let $\tau_{B}$ be the average sampling time of the field distribution. 
As long as $\tau_B \gg \tau_{sys}$ 
the spin system has enough time to
relax to global thermal equilibrium. 
This is the case of equilibrium
statistical mechanics and normal fluctuations. 
The situation can be very different if the field can
switch abruptly ($\tau_{switch}^{-1} \sim \mu_B \dot B$ is large) 
and $\tau_{sys} \gg \tau_B $. The system does
not have enough time to relax to equilibrium and the stationary
state is determined by the external random field. As discussed later, 
such a situation is experimentally realizable.

Now let us assume that the field is sampled from $\rho(B)$ at
time intervals of length $\tau_B$,
\begin{equation}
    B(t) = \mu_B B \rho(B) \sum_{n=0}^{\infty}\Theta(t-n \tau_B)\Theta \left((n+1)\tau_B - t\right)
    \label{h(t)}
\end{equation}
We can integrate Eq. (\ref{mastereq}) exactly
from $t_{n-1} = (n-1)\tau_B$ to $t$  for $t_{n-1} < t < t_n$:
\begin{equation}
       | P(\{ s_i \};t) \rangle = {\rm e}^{- \hat {\cal L}_{B(t_{n-1})} (t-t_{n-1})} | P(\{ s_i \};t_{n-1}) \rangle 
        \label{evolop1}
\end{equation}
where $B(t_{n-1})$ is the field instance sampled at time $t_{n-1}$. 
For $t = \lim_{\epsilon\to 0} (t_n - \epsilon)$ one obtains 
\begin{equation}
       | P(\{ s_i \};t_n) \rangle  = {\rm e}^{- \hat {\cal L}_{B(t_{n-1})} \tau_B } | P(\{ s_i \};t_{n-1}) \rangle  
        \label{evolop2}
\end{equation}
Since in our case the low 
lying eigenvalues $\lambda_{\alpha}$ of the Liouville operator $\hat{\cal L}$ satisfy $\lambda_{\alpha} \tau_B \ll 1$, 
by expanding the exponential in first order one obtains 
the discrete `coarse grained' Master Equation
\begin{equation}
        { | P(\{s_i\};t_n)\rangle -| P(\{s_i\};t_{n-1})\rangle \over \tau_B}  = - \hat {\cal L}_{B(t_{n-1})} | P(\{s_i\};t_{n-1})\rangle       
        \label{cgmastereq}
\end{equation}
which describes correctly only the long-time behavior of Eq. (\ref{mastereq}). Short term
effects due to the larger eigenvalues of $\hat {\cal L}$ have already relaxed at the time scale $\tau_B$.
This approximated form of the Master Equation is used in all further developments.

One can regard Equation (\ref{cgmastereq}) as defining the discrete dynamics
governing the probability distribution $|P(\{s_i\};t_n)\rangle$. 

\subsection{The driving field distribution}

As already mentioned, the external field might be distributed according to
the invariant measure of some chaotic one-dimensional deterministic map.
In other applications, the field can be Poisson- or Gauss-distributed. 
In what follows we restrict ourselves to the binary distribution
\begin{equation}
        \rho(B) = {1\over 2} \delta(B - B_0) + {1\over 2} \delta(B + B_0)
        \label{rhoH}
\end{equation}

Many of our results can easily be
generalized to arbitrary continuous distributions. Other results, in particular
those concerning the stationary state phase diagram and the critical behavior, 
depend strongly on the 
discrete character of the choice (\ref{rhoH}).
Eqs. (\ref{cgmastereq}) and (\ref{rhoH}) 
map our problem into an iterated function system (IFS) \cite{IRF}. As long as we deal
with a finite system of spins, the mathematical results (including the collage theorem)
developed by Demko and Barnsley and subsequent work on IFS apply to randomly
driven spin models as well. However, from a statistical physical point of view, the
interesting things happen {\it after} taking the thermodynamic limit.

\subsection{The invariant measure}

As usual in the theory of dynamic systems, one can ask what is the invariant measure
induced by the dynamics (\ref{cgmastereq}), see, for example, \cite{wright}. 
Let ${\cal P}_s (\vec \pi)$ denote the invariant density related to the dynamics
Eq. \ref{cgmastereq}.
It satisfies the Chapman-Kolmogorov equation:
\begin{equation} 
        \label{CK}
        {\cal P}_s (\vec \pi)  = \int  d \vec \pi' \ {\cal P}_s(\vec \pi') \int dB \ \rho(B) \
                                \delta(\vec \pi - {\rm e}^{- \hat {\cal L}_{B} \tau_B}\ \vec \pi')
                                 \equiv  \tilde {\cal K} {\cal P}_s (\vec \pi)
\end{equation}
where $\tilde {\cal K}$ denotes the Frobenius-Perron (FP) operator. 
Physically, ${\cal P}_s (\vec \pi)$
describes the nonequilibrium stationary state induced by the Master Equation dynamics. 
Note that we used above the spin-product basis. An orthonormal transformation of the basis will lead to a different but equivalent FP-operator, 
the transformation's Jacobian is unity. Again, this is 
true only for finite systems.

If one is interested in the stationary expectation
value of some spin observable $A(\{s_i\})$ one must perform two averages, 
the `thermal' $\langle ...\rangle$
and the `dynamic' average $[...]$:
\begin{equation}        
        \overline{A(\{s_i\})} =  [ \langle A \rangle ]
        \label{aver}
\end{equation}
where the `dynamic' average $[...]$ is taken over ${\cal P}_s$ {\it and} $\rho(B)$. 

The `thermodynamics' of such driven systems can be computed from the generalized
free energy. This is related to the largest Lyapunov exponent $\Lambda$ of the dynamics
as $-\beta {\cal F} = \Lambda$. Consider a long dynamic trajectory consisting of $T\tau_B$
sampling points. The Lyapunov exponent is defined as
\begin{equation}
        \Lambda = \lim_{T \to \infty} {1\over \tau_B T} \ln {\rm Tr} \left\{ {\rm e}^{- \hat{\cal L}_{B(T)} \tau_B} {\rm e}^{- \hat {\cal L}_{B(T-1)}\tau_B}\cdots {\rm e}^{- \hat{\cal L}_{B(1)}\tau_B} \right\}  
        \label{lyap1}
\end{equation}
where $B(n)=B(t_n)$ is distributed according to Eq. (\ref{rhoH}). The same result can be obtained
by iterating some general\footnote{This vector should not fall into any invariant subspace of
{\it both} $ \hat {\cal L}_{\pm B_0}$ operators} initial unity vector $|p_0 \rangle$ as
\begin{eqnarray}
        | p_1\rangle \  & = & \ {\rm e}^{-\hat{\cal L}_{B(1)}\tau_B} | p_0 \rangle, \ a_1  = \sqrt{ \langle p_1 | p_1\rangle } \nonumber \cr
        | p_2\rangle & = & {1\over a_1} {\rm e}^{-\hat {\cal L}_{B(2)}\tau_B} | p_1 \rangle, \ a_2  = \sqrt{ \langle p_2 | p_2\rangle } \nonumber \cr
        \dots & & \nonumber 
\end{eqnarray}
For large $n$ the vectors $|p_n\rangle$ will be distributed according to ${\cal P}_s$ and
up to $O({1\over T})$ corrections the Lyapunov exponent can be expressed as
\begin{equation}
        \Lambda = \int  d{\cal P}_s(\vec \pi) \int dB \ \rho(B) {1\over 2}  \ln  \Vert {\rm e}^{-\hat {\cal L}_B \tau_B} {\vec \pi} \Vert
        \label{freeen}
\end{equation}
For a constant field $\rho(B) = \delta(B-B_0)$ and the stationary distribution is ${\cal P}_s(\vec \pi) =
\delta(\vec \pi - \vec \pi_{eq}) $, where $\vec \pi_{eq}$ are the 
Boltzmann-distribution averaged spin-products. 
Therefore, $\Vert {\rm e}^{-\hat {\cal L}_{B_0} \tau_B} {\vec \pi}_{eq} \Vert = 
(\langle l_0 | r_0 \rangle)^2 = Z^2$, where we have used that $\vec \pi_{eq}$ is the right eigenvector of $\hat {\cal L}_{B_0}$ with eigenvalue 0. We recover the usual definition of free energy by multiplying $\Lambda$ with $-k_B T$.

This generalized free energy might display
anomalous fluctuations related to the multifractal spectrum of the 
stationary distribution ${\cal P}_s$, as will be shown later for the mean field theory.

\subsection{Dynamical properties}

In order to consider the {\it dynamical} properties of randomly driven systems 
one has to solve - in full analogy to the theory of one-dimensional maps - 
the right eigenvalue problem of the
Frobenius-Perron operator:

\begin{equation}
        \label{FPeigv}
        \tilde {\cal K} {\cal R}_m = s_m {\cal R}_m
\end{equation}
The largest magnitude eigenvalue is one, $s_0 = 1 $.
The right eigenvector ${\cal R}_0$ is nodeless and real:
it corresponds to the stationary state,  ${\cal R}_0 = {\cal P}_s$. 
For $m>0$ the eigenvectors $R_m(x)$ satisfy
\begin{equation}
        \int d\vec \pi\ R_m(\vec \pi) = 0
        \label{intrxeqnull}
\end{equation}
In addition to the right eigenvectors ${\cal R}_m$, the operator $\tilde {\cal K}$
might have also a set of null functions of different orders. They
correspond to zero eigenvalues of ${\tilde {\cal K}}^q$, where $q$ is an integer
(the order). These eigenfunctions, however, do not contribute to the relaxation
of the initial probability distribution towards ${\cal P}_s$.

In analogy to the usual transfer matrix theory, the relaxation of the
probability distribution and of the time dependent 
correlation functions are determined for asymptotically long times by 
the second largest eigenvalue $s_1$ and its eigenfunction ${\cal R}_1$.

\vspace{1cm}

Although at this stage the formalism looks rather involved, it is a straightforward extension
of the methods developed for low-dimensional dynamic systems. 
We consider next
the simplest possible example, an
Ising model in a random binary external field, 
Eq. (\ref{rhoH}). In this case, many interesting
stationary and dynamic properties can be obtained analytically, 
or with numerical methods not more complex than those used for one dimensional maps.


\section{Mean Field Approximation}
\subsection{The Mean-Field Map}

Consider an Ising model defined on an
$N$-dimensional simplex, such that all spins are nearest neighbors:
\begin{equation}
        \label{mfenergy}
        E = - {J\over N} \sum_{i\neq j} s_i s_j - \mu_B B \sum_i s_i
\end{equation}
$J$ is normalized so that the energy is additive and $\mu_B$ is the Bohr magneton. 

Let ${\vec \mu}_i := (s_1, \dots,-s_i, \dots,s_N)$. We may describe the Liouville
operator $\hat {\cal L}$ with the transition rate $w(\vec \mu_i|\vec \mu)$ in the
Glauber form\cite{glauber}
\begin{equation} 
        \label{trprob}
        w(\vec \mu_i|\vec \mu) = {1\over {2 \alpha}} [1 - s_i {\rm tanh}({ K\over N} \sum_{j\neq i} s_j + H)]
\end{equation}
where $\beta = 1/k_BT$, $K=\beta J$, $H=\beta \mu_B B$ and $\alpha$ sets the time constant.
Applying Eq. (\ref{cgmastereq}) one obtains after performing the thermodynamic limit $N\to \infty$:
\begin{equation}
        m(t+1)  = {\rm tanh}(Km(t) + H(t)) ,
        \label{mfmap} 
\end{equation}
Time is measured in units of $\tau_B$. The field distribution Eq. (\ref{rhoH}) leads
to the one-dimensional map
\begin{equation}
        \label{mfbimap}
    m(t+1) = \left\{ \begin{array}{lcl} 
                      {{\rm tanh}(Km(t) + H_0)}\ & {\rm with\ probability} &\ {1\over 2} \\
                     & & \\
                      {{\rm tanh}(Km(t) - H_0)}\ & {\rm with\ probability} &\ {1\over 2} \\
                        \end{array} \right.
\end{equation}

Note that in the thermodynamic limit
the moments of the magnetization do not couple with higher order correlation 
functions and the methods
worked out previously for the one-dimensional random-field Ising chain can
thus be applied directly.

Since in the stationary state, Eq. (\ref{CK}), $[m^k(t+1)]=[m^k(t)]$, 
using Eq. (\ref{mfmap}) and simple algebraic manipulations 
we obtain that the $k$-th moment of the stationary magnetization is given by
\begin{equation}
        \label{mfmoments}
        [m^k] = \left[ \left({v+h \over 1 + v h}\right)^k \right]\ \ \ \ \ k=1,2,...
\end{equation}
where $v={\rm tanh}(K m)$ and  $h={\rm tanh}(H)$.

At high temperature the system is in the disordered, paramagnetic phase,
in which case all odd moments of the magnetization vanish. {\it Assuming} that
the free energy is analytic in $[m]$, the critical temperature is
obtained by expanding $[m]$ in first order in O$([h^2])$:
\begin{equation}
        \label{mom1}
        [m ]  \simeq K(1-h_0^2)  [m] 
\end{equation}
and neglecting $[m^q],\ q=3$ and higher odd moments (which should scale as O$([h^{2q}])$).
In the usual mean-field scenario $[m]=0$ in the paramagnetic phase and the
coefficient vanishes at the transition point to the ferromagnetic phase:
\begin{equation}
        \label{pht-cont}
        H_c^{(II)} = {1\over 2 } {\rm ln}{1+m^{\dag}\over 1-m^{\dag}} ,
\end{equation}
where $m^{\dag} = \pm \sqrt{K-1\over K}$ for $K > 1$. 
Using a simple geometric argument we will show below that this
analyticity assumption fails and the phase transition is actually first order.
For the second moment one obtains 
\begin{equation}
        [m^2]  \simeq  {h^2_0 \over  1 - K^2 (1-4h^2_0+3h^4_0) } 
        \label{mom2} 
\end{equation}
where we have omitted O$([m^4])$ and higher even moments. The pole of this
expression is also related to the phase transition, which is discussed below.
Third and fourth order expansions of $[m]$ and $[m^2]$ read
\begin{equation}
        \label{mom3}
        [m ]  \simeq \left( K(1-h_0^2) - \frac{K^4(h_0^2 -h_0^4)}{1 - 
            K^3(1 - 10 h_0^2 + 19 h_0^4 -10 h_0^6)} \right) [m] 
\end{equation}
and
\begin{multline}
        \label{mom4}
        [m^2 ] \simeq \left( h_0^2 - \frac{h_0^4 K^4 (2 - 17 h_0^2 + 30 h_0^4 - 15 h_0^6)}
          {3 - K^4(3 - 60 h_0^2 + 212 h_0^4 - 260 h_0^6 + 105 h_0^8)}\right) \times \\
        \left( 1 - K^2(1 - 4 h_0^2 + 3 h_0^4) +  \frac{K^6 ( 2 -  17 h_0^2 + 30 h_0^4 - 15 h_0^6)
            (6 h_0^2 - 16 h_0^4 + 10 h_0^6)}{3 - K^4(3 - 60 h_0^2 + 212 h_0^4 - 260 h_0^6 +
            105 h_0^8)}\right)^{-1}
\end{multline}
respectively. A high order expansion of the moments along these lines
can be easily obtained using algebraic manipulations programs but will not be
presented here.

\subsection{The stationary phase diagram}

In principle, there are at least two different mechanisms for a phase transitions in the
stationary state described by ${\cal P}_s(m)$. 
The first one corresponds to spontaneous symmetry breaking
leading to a continuous phase transition. 
In this scenario  the stationary distribution, 
which at high temperature is a function of the
even magnetization moments only, ${\cal P}_s(m) = {\cal P}_s(-m)$, 
becomes degenerate at certain parameter values
$\{K,H_0\}$ and the odd subspace, ${\cal P}_o(m) = -{\cal P}_o(-m)$,
contributes as well. More precisely, 
${\cal P}_s(m) = {\cal R}_0$
is {\bf always} a nodeless even function of $m$. 
The field symmetry is spontaneously broken when   
the largest odd-subspace eigenvalue of the Frobenius-Perron operator Eq. (\ref{FPeigv}),
$s_1 \to 1$. Therefore, the largest eigenvalue is degenerate and the
corresponding eigenvector is an arbitrary linear combination of ${\cal P}_s(m)$ and
${\cal R}_1(m) = -{\cal R}_1(-m)$, leading to a non-vanishing order parameter.
Close to but above the transition point the relaxation time of the stationary
distribution diverges as ${\tau}^{-1} \sim 1-s_1$. We find no evidence for such
a mechanism, at least not in mean-field approximation. 
Instead, the phase transition is related to a 
bifurcation of the stationary magnetization distribution. 

\begin{figure}
  \begin{center}
    \leavevmode
    \epsfysize=6.0truecm \epsfbox{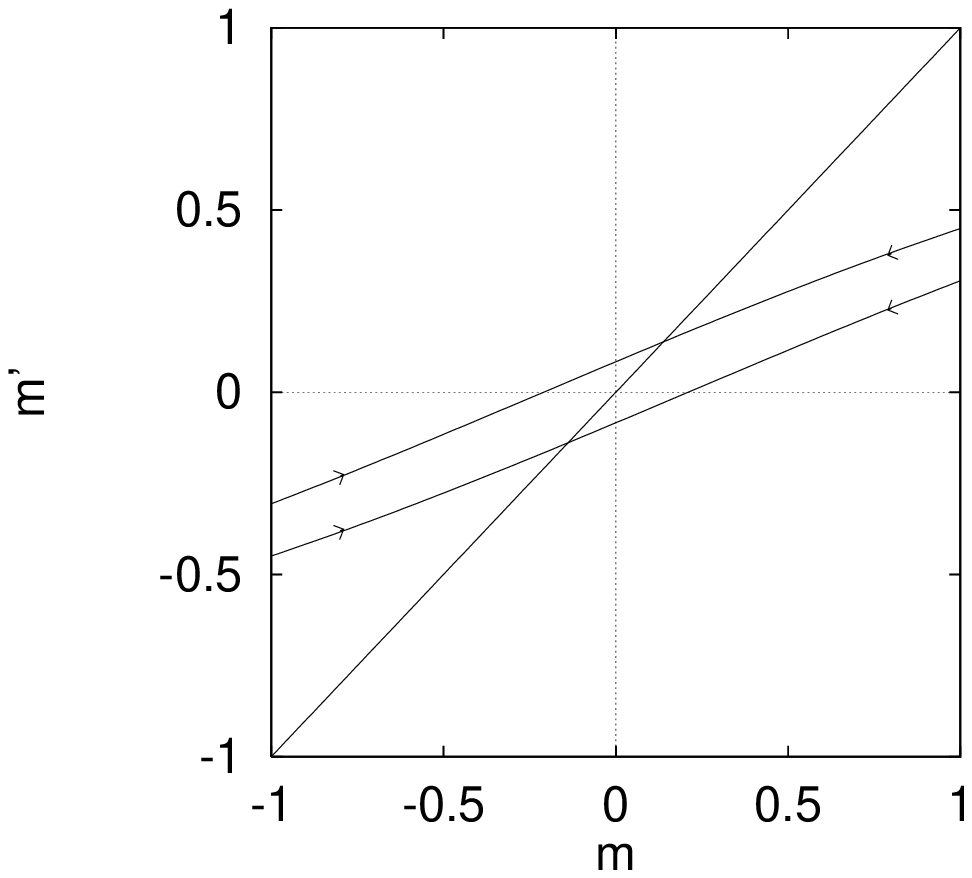}
    \epsfysize=6.0truecm \epsfbox{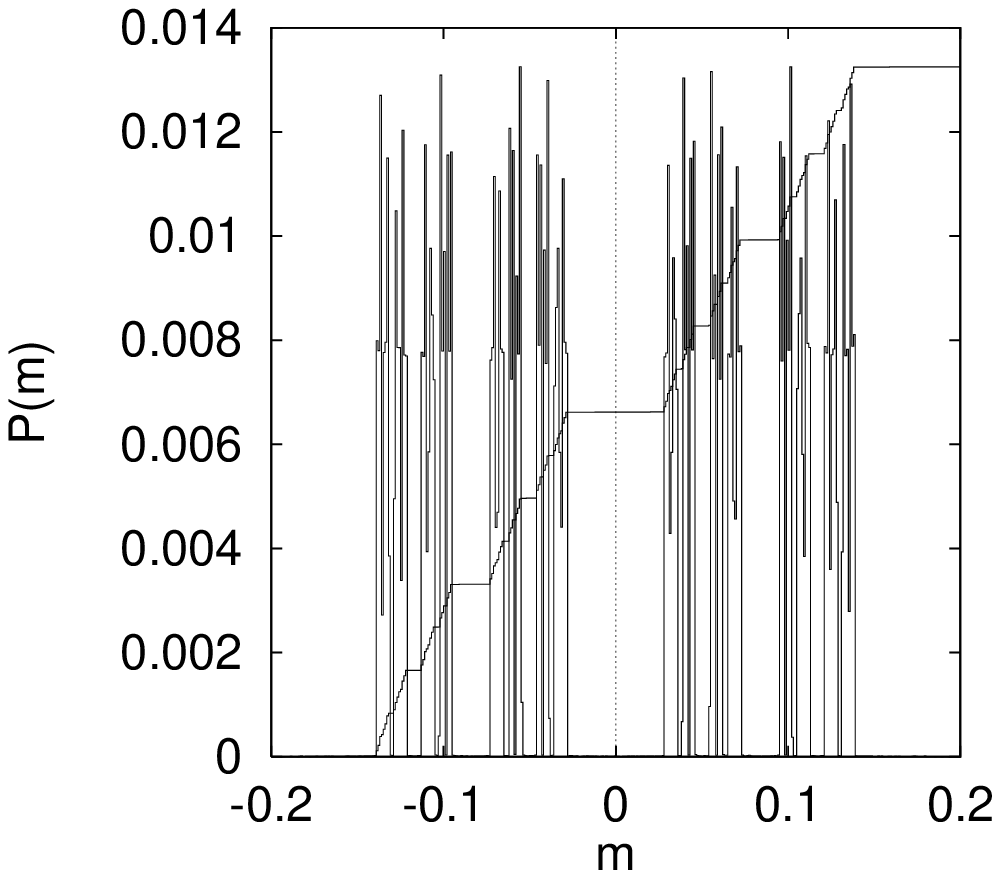}
  \end{center}
  \caption{\it Mean field map and the stationary 
    distribution in the paramagnetic phase ($K = 0.4$ and $H_0/K = 0.21$).
    }
  \label{snap0}
\end{figure}

\begin{figure}
  \begin{center}
    \leavevmode
    \epsfysize=6.0truecm \epsfbox{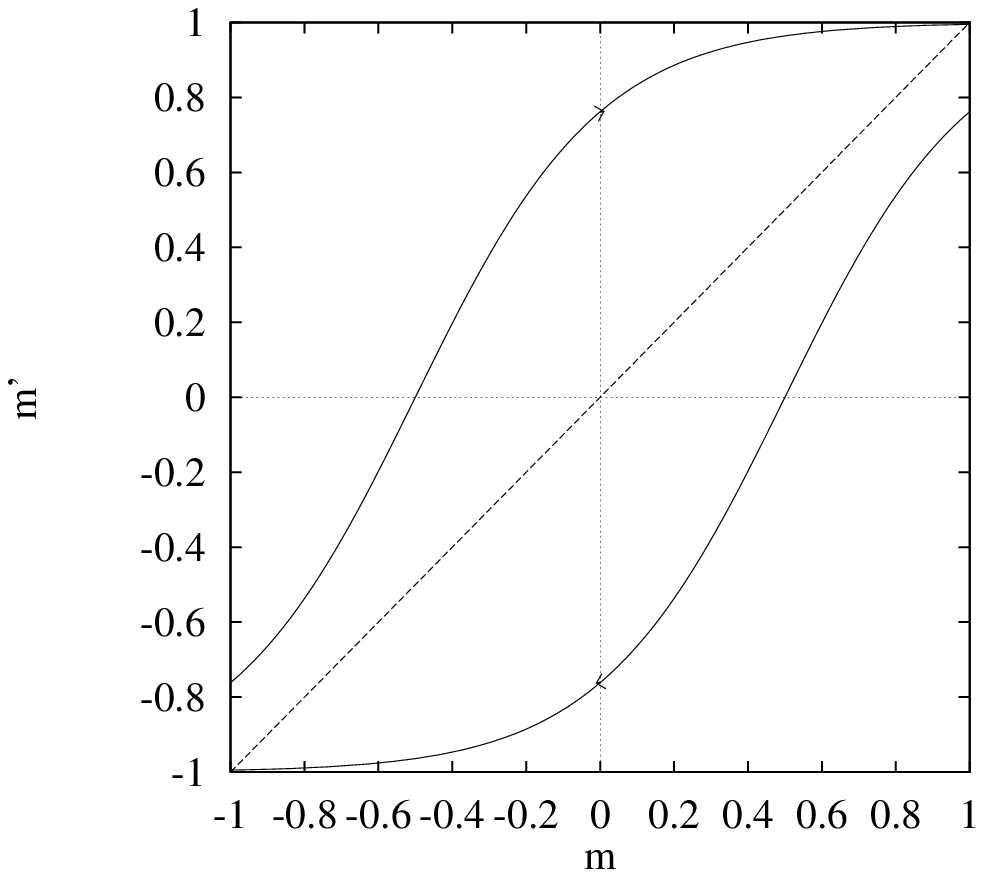}
    \epsfysize=6.0truecm \epsfbox{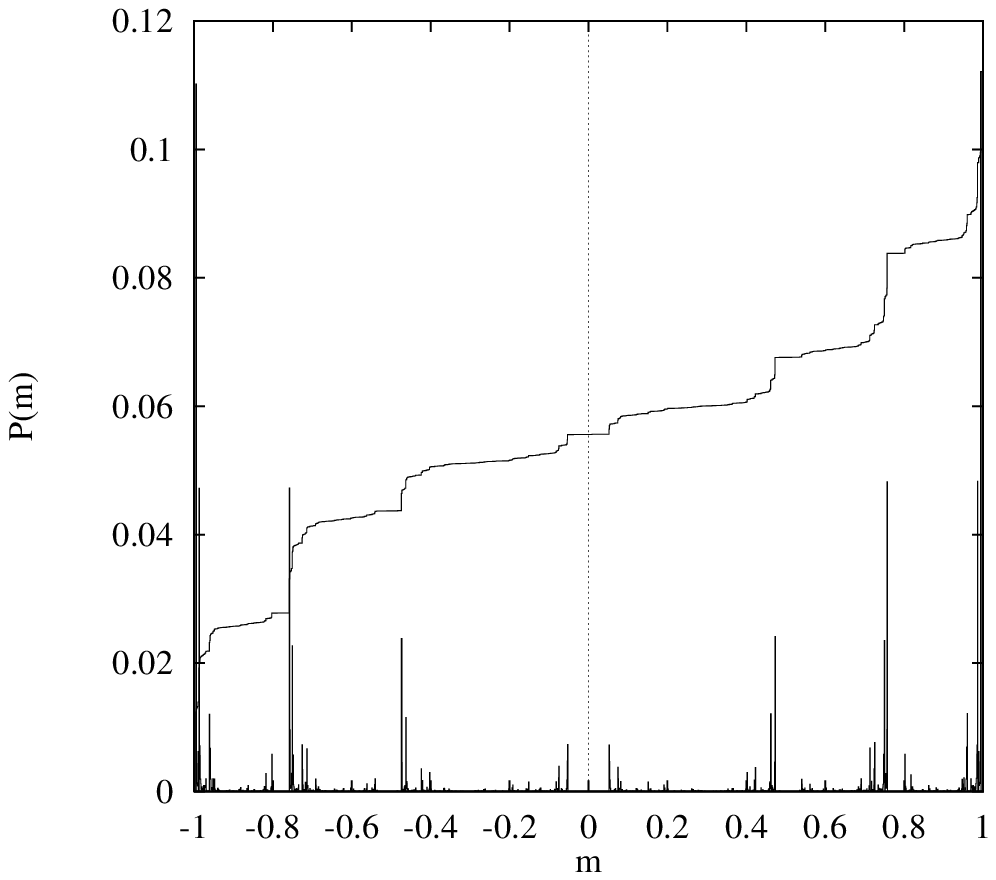}
  \end{center}
  \caption{\it Mean field map and the stationary 
    distribution for $K = 2.0$ and $H_0/K = 0.5$. The map shows that
    for strong driving fields the system remains in the paramagnetic phase
    even below the equilibrium critical temperature.
    }
  \label{snap1}
\end{figure}

\begin{figure}
  \begin{center}
    \leavevmode
    \epsfysize=6.0truecm \epsfbox{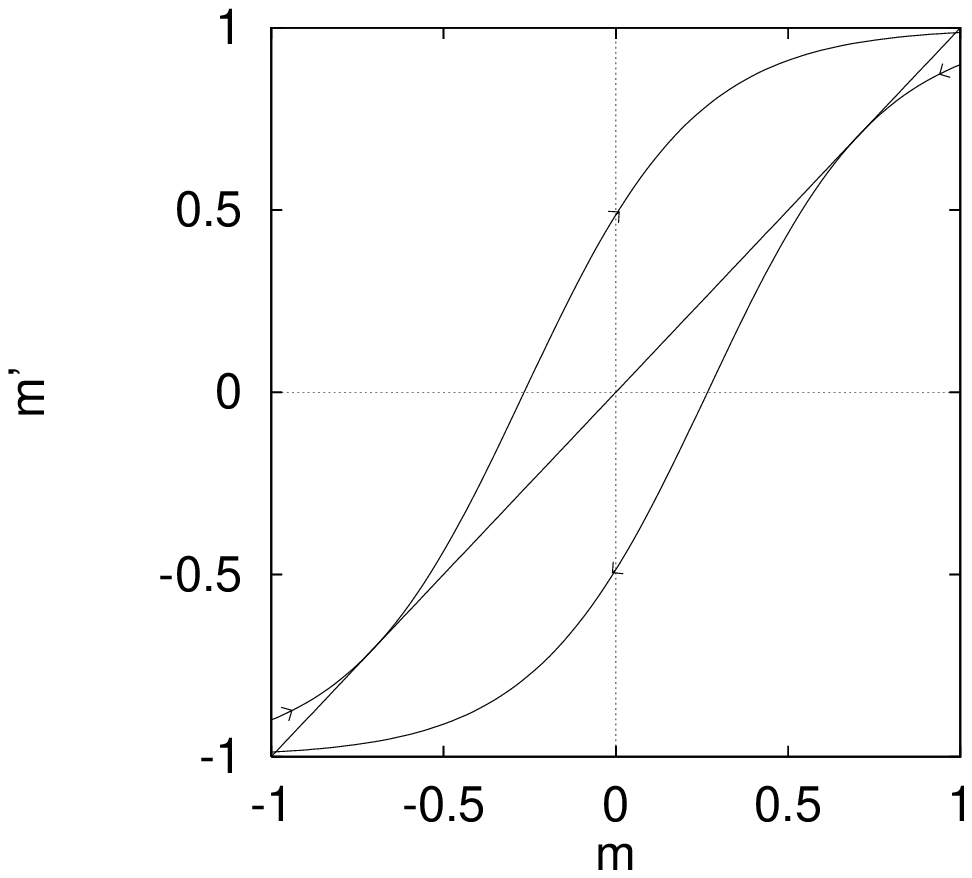}
    \epsfysize=6.0truecm \epsfbox{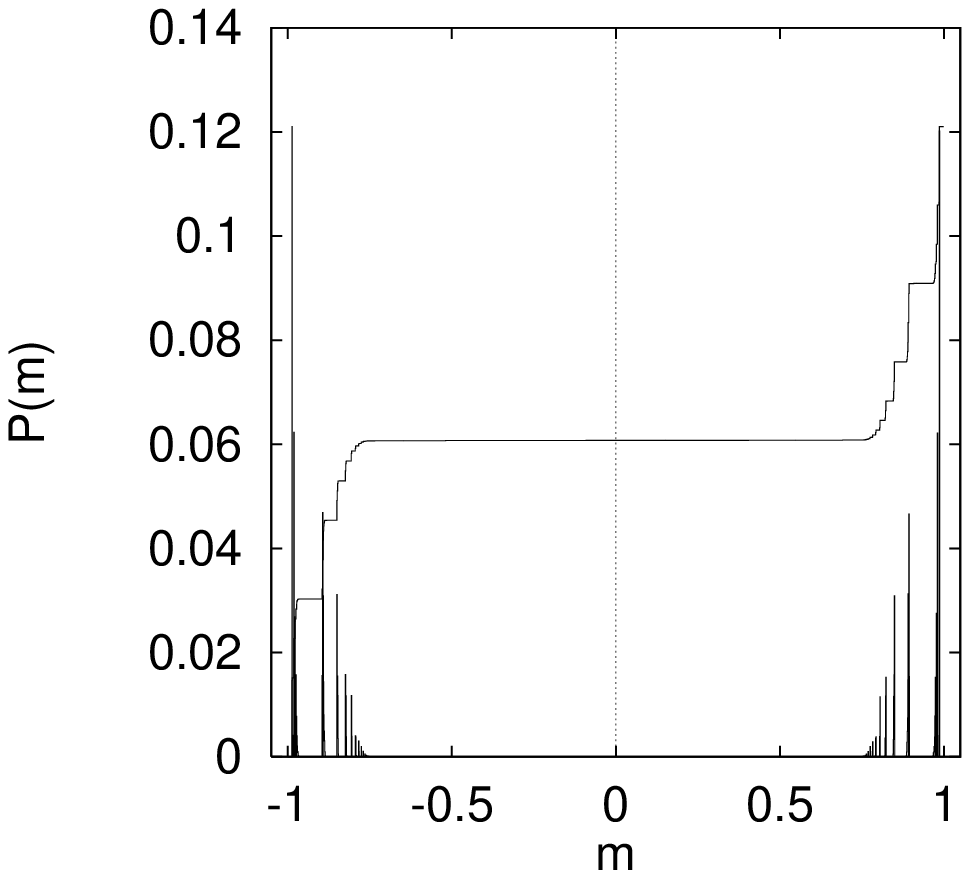}
  \end{center}
  \caption{\it Same as in Fig. 1 but close to the
    critical field value ($K = 2.0$ and $H_0/K = 0.266$).
    Two disjoint distributions are created around the stable fixed points,
    a repellor in the middle.}
  \label{snap2}
\end{figure}

\begin{figure}
  \begin{center}
    \leavevmode
    \epsfysize=6.0truecm \epsfbox{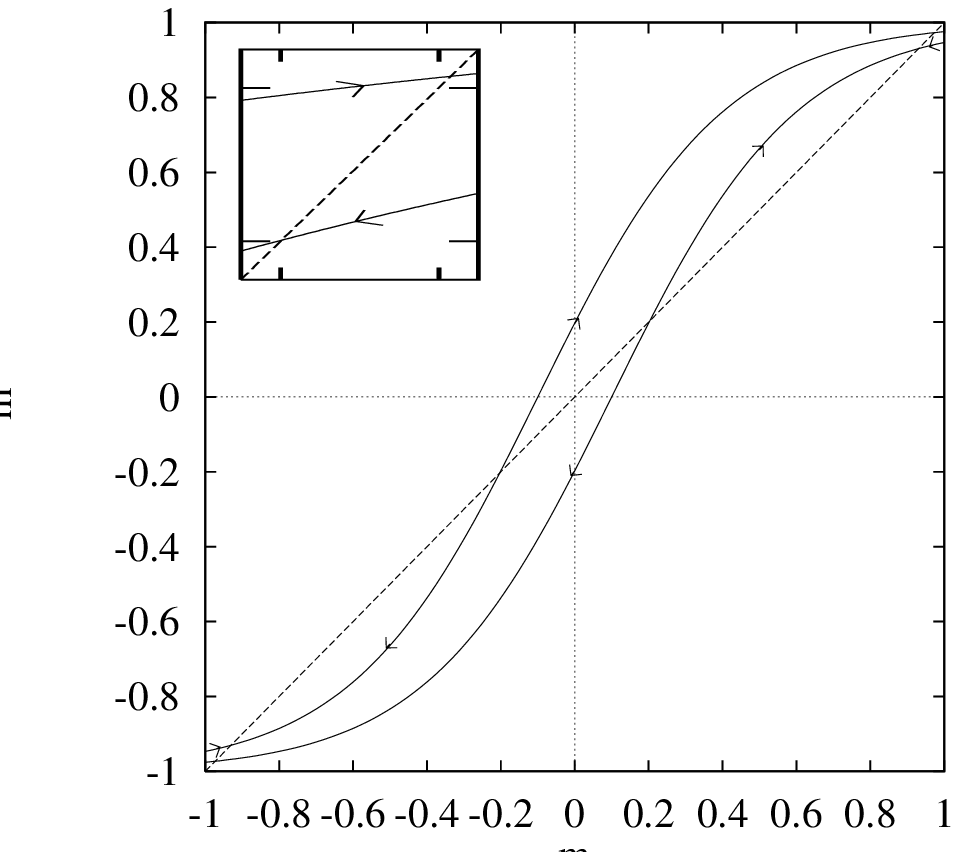}
    \epsfysize=6.0truecm \epsfbox{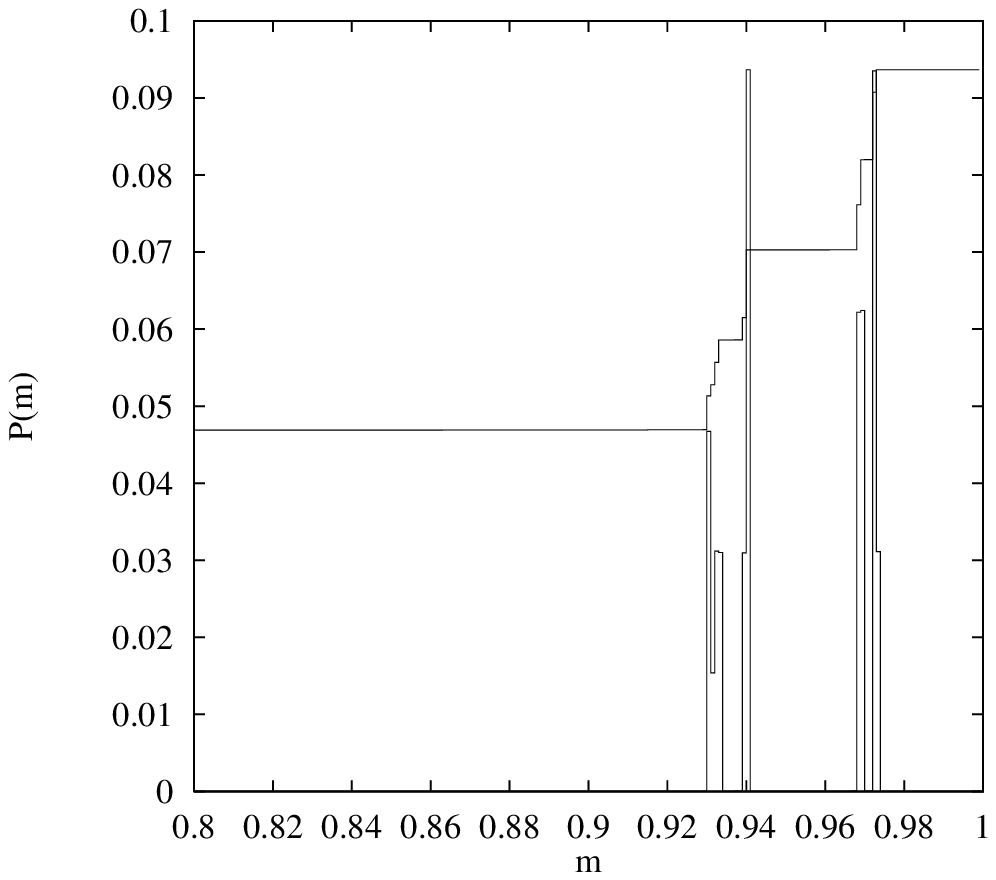}
  \end{center}
  \caption{\it Mean field map and the stationary 
    distribution in the ferromagnetic phase ($K = 2.0$ and $H_0/K = 0.1$).
    }
  \label{snap3}
\end{figure}

Consider the map Eq. (\ref{mfbimap}) at high temperature, a situation shown in 
Fig. \ref{snap0}.
The arrows indicate the direction of the flow. The competition between the two stable
fixed points leads to chaotic behavior and the displayed stationary distribution. To
approximate the distribution, we tracked the evolution of 1000 (random) initial values
of $m$ subject to the map for 1000 iterations.
At low temperatures ($K > 1$) and large fields one has the situation depicted 
in Fig. \ref{snap1}.
Note the possible intermittent behavior close to $m\sim 0$. If we decrease $H_0$
the map can `pinch' tangentially the $m(t+1) = m(t)$ diagonal, 
creating thus one new unstable fixed point. This situation is shown in Fig. \ref{snap2}. 
Decreasing the field even further, we
have the map of Fig. \ref{snap3}, where ${\cal P}_s$ has bifurcated into two stable and one
unstable disjoint distributions.
For further use let us denote by 
$m_1,\ m_2$, and $m_3$ the possible
fixed points of the equation $m=\tanh(Km+H_0)$ in descending order.
The line of the critical field $H_c$ can be calculated from
the condition that at the new fixed point the map is tangential (`critical map', see \cite{mp})
and leads after elementary calculations to 
\begin{equation}
        \label{pht-first}
        H_c = {1\over 2 } {\rm ln}{1-m^{\dag}\over 1+m^{\dag}} + Km^{\dag}
\end{equation}
where $m^{\dag}=m_2 = m_3$.

\begin{figure}
  \begin{center}
    \leavevmode
    \epsfysize=6.0truecm \epsfbox{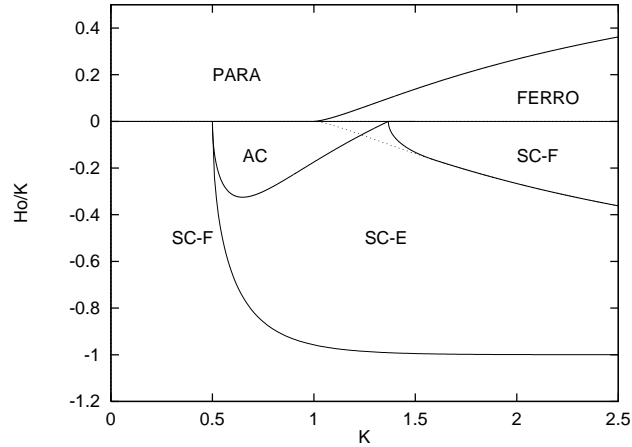}
  \end{center}
  \caption{\it  Mean field phase diagram. The upper part $(H_0>0)$ shows the
    border between the para- and ferromagnetic phase. In the lower part
    $(H_0<0)$ the regions
    denoted by SC-F, SC-E correspond to  
    a singular-continuous invariant density 
    with fractal and Euclidean support, respectively,
    while in the AC-region the density is absolutely-continuous. 
    Note that the diagram is actually symmetric in $H_0$.}
  \label{mfphgddiag}
\end{figure}

From Figs. \ref{snap1}-\ref{snap3} it is evident that $ |m| > m^{\dag}$ 
and except for $H_0=0$ the magnetization jumps at the phase transition. 
We believe that this feature is due to the discrete character of the binary 
$\rho(B)$ distribution. Thus, the RDIM provides an example of a spontaneous 
symmetry breaking leading to a first order phase transition. 
The mechanism behind this first
order transition is very different from that of equilibrium systems and is
related to a tangential bifurcation of the stationary distribution.
The corresponding phase diagram
is shown in the upper part of Fig. \ref{mfphgddiag}. 

\subsection{The multifractal regime}
Some highly unusual properties of the RDIM are related to the multifractal
spectrum of the stationary state.
Following the notation introduced in \cite{radons}, one can identify
a singular-continuous density with fractal support (SC-F) in both the
paramagnetic and the ferromagnetic phase. 
When a gap opens between the upper and the lower branch of the map the 
invariant distribution has a fractal support with the capacity
dimension $d_0 < 1$. The border of the (SC-F) region is given by $Km_1 = H_0$ in the
para- and $K(m_1+m_3)=2H_0$ in the ferromagnetic phase.  In the
region between $d_0 = 1$ and $d_{\infty}=1_-$ the distribution is
singular-continuous with Euclidean support (SC-E) \cite{radons}. Using the ideas developed in
\cite{evang}, we obtain $d_{\infty}=1$ if $K(1-m_1^2)={1\over2}$.
The density distribution is absolutely continuous (AC) 
if all generalized dimensions \cite{hp} equal one, $d_q=1$, $(q=0,\dots,\infty)$. 
These results are graphically summarized in the lower part of Fig. \ref{mfphgddiag}. 

In order to compute the generalized free energy, Eq. (\ref{freeen}), 
one can use that 
$\langle l_0 | r_n \rangle = \delta_{0,n} \sum_{\{s_i\}} P_{eq}(\{s_i\}) $. Note that
the left eigenvector $\langle l_0 |$ of the $\hat {\cal L}_B$ operator 
corresponds to a sum over all spin configurations
and is therefore independent of $B$. When inserting in the product within the
trace of Eq. (\ref{lyap1}) the spectral decompositions of different non-commuting operators 
$\hat {\cal L}_B$, the ground state contributions decouple from the higher level contributions. 
Hence, in mean field approximation the free energy is given as expected by
\begin{equation}
        \label{mffe}
        -\beta {\cal F} = N \int dm {\cal P}_s(m) {1\over 2} \ln 2 [\cosh(2 K m) + \cosh (2 H_0)]
\end{equation}
This integral can be approximated above $H_c$  by expanding the integrand in even moments
of the magnetization (see Eq. (\ref{mom2}) and (\ref{mom4})). The fluctuations of the free
energy depend on the (multifractal) structure of the stationary ${\cal P}_s(m)$ distribution.

Strictly speaking, Eq. (\ref{mffe}) is the average free energy. When considering a finite system or
a long but finite dynamic trajectory, the free energy is normally distributed. As shown in \cite{peter2} 
for the one dimensional random field Ising model,
in the SC-F region the multifractal spectrum can be directly related to the second cumulant of the 
free energy distribution. The arguments presented in \cite{peter2} apply also to our case, 
a broad multifractal distribution leads to large free energy fluctuations.

In the SC-F regime one can obtain additional information about finite-size
free-energy fluctuations from the generalized dimensions (Legendre-transform
of the multifractal spectrum). 
We applied
the methods developed in \cite{evang,peter1,peter2,bene,tsang} 
and computed numerically the multifractal
spectrum of the stationary distribution. 

Another interesting observation is that these isolines cannot 
directly cross into the ferromagnetic region:
close but above the phase transition there is no positive gap 
(see Figs. \ref{snap1}--\ref{snap2}).
Nevertheless, in the ferromagnetic phase the magnetization
distribution itself can be multifractal. 
This is shown in Fig \ref{snap3}. The inset shows the enlarged part of the
map leading to a multifractal distribution for positive magnetization (a symmetric
counterpart exists for negative magnetization).

\subsection{Dynamical properties}

\begin{figure}[htbp]
  \begin{center}
    \leavevmode
    \epsfysize=6.0truecm \epsfbox{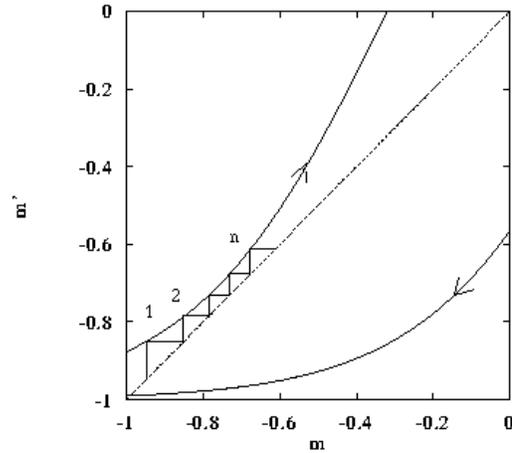}
    \caption{\it Lower part of the mean field map for $H \geq H_c$. Iteration along the upper branch is type I intermittent while the lower branch brings the iteration to the starting point in one step.}
    \label{snap2me}
  \end{center}
\end{figure}

The stationary phase transition at $H_c$, Eq. (\ref{pht-first}), is from a physical
point of view a {\bf dynamic freezing} transition characterized by an extremely slow
dynamics. As shown below, the relaxation of the map - and hence of all time-dependent
correlation functions - diverges exponentially fast close to the critical
field $H_c$.  Consider first the mean-field map close but above the critical
field, as illustrated in Fig. \ref{snap2me}.

The iteration along the upper branch alone corresponds to type I 
intermittency and has been discussed previously
in the theory of chaotic maps \cite{mp,gh,gsz}. As usual, the function
$m' = \tanh(Km+H_0)$ will be approximated up to quadratic order close 
to the point $(m^{\dag},H_c)$ where $m^{\dag} = \pm \sqrt{K-1\over K}$
is the point where the upper branch touches tangentially the $m'= m$ line.
Introducing the new variable $x = {m - m^{\dag} \over K(1 - {m^{\dag}}^2)}$, 
one obtains
\begin{equation}
        \label{interm}
        x_{n+1} =  x_n + m^{\dag} {x_n}^2 + {H_0 - H_c \over K}
\end{equation}
Requiring that $x_{n+1} - x_n \over \delta n$, $x^2$, and ${H_0 - H_c \over K}$
have the same order of magnitude implies that $\delta n$ and hence $n$
must scale as 
\begin{equation}
        \label{sqrt}
        n \sim \left({H_0 - H_c \over K}\right)^{-{1\over 2}}
\end{equation}
which is the standard result for one-dimensional maps \cite{mp}.

\begin{figure}[htbp]
  \begin{center}
    \leavevmode
    \epsfysize=4.0truecm \epsfbox{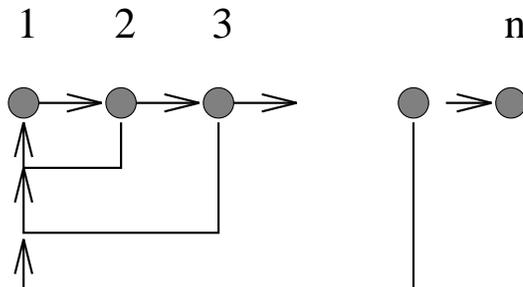}
    \caption{\it The Markov process representing the passage through an intermittent tunnel. Each arrow indicates a transition probability of $1 \over 2$.}
    \label{int-markov}
  \end{center}
\end{figure}

However, the probability to stay on the upper branch of the map for $n$ consecutive
steps is exponentially small. Assume that at time
$t=0$ one injects $N_0$ points at the $m = -1$ location. 
In order to move upwards, the points can use only the upper branch and must
pass through the `intermittent tunnel'. Once a trajectory flips
to the lower branch, it is set back to the entrance of the tunnel.
If a point has passed through the tunnel, it might eventually return
to the lower part but has a similar chance of being trapped on the
symmetric upper part. This dynamics can be modeled by the 
Markov process shown in Fig. \ref{int-markov}.

By iterating the corresponding stochastic matrix (or by full induction)
it is easy to see that
the stationary probability of being at site $n$ is given by $p_n = {1\over 2^n}$.
Therefore, assuming quasi equilibrium, the escape rate is estimated as
\begin{equation}
        \label{escrate}
        \dot N(t) = - {a\over 2^n} N(t)
\end{equation}
where $a$ is a constant of order O(1) related to the probability of
return after escape. The relaxation time $\tau$ corresponding to Eq. (\ref{escrate})
diverges as
\begin{equation}
        \label{relexp}
        \tau = {2^n \over a} \sim 2^{\alpha [H_0-H_c]^{-{1\over 2}} }
\end{equation}
where we have used Eq. (\ref{sqrt}), $\alpha$ is a constant. 
Hence, the relaxation time diverges 
exponentially fast close to the phase transition.
Below $H_c$, the slow dynamics is due to the
average escape time (fractal dimension) from the central repellor.

\begin{figure}[htbp]
  \begin{center}
    \epsfysize=6.0truecm \epsfbox{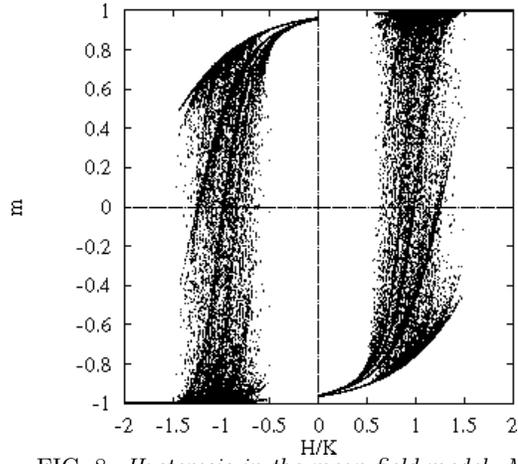}
    \caption{\it Hysteresis in the mean field model. Magnetization $m(t)$ vs. the external driving field $H(t)$, see Eq.~(\ref{hystfield}), for parameters $\frac{A}{K} = \frac{H_0}{K} = 1$, $K=2$, and $\Omega = \frac{2 \pi}{1000}$.}
    \label{mfhyst}
  \end{center}
\end{figure}

Another interesting dynamic phenomenon is the hysteresis of the RDIM.
Here one adds a harmonic part to the external driving field:
\begin{equation}
        H(t) \mapsto H(t) + A \cos \Omega t
        \label{hystfield}
\end{equation}
The resulting hysteresis distribution is shown in Fig.~\ref{mfhyst}. The evolution
of 500 initial values was tracked for 2000 iterations.
We close here the discussion of the mean-field (or infinite dimensional) RDIM. 
We expect many of
the features discussed here to be valid in three dimensional systems and to
a lesser extent in two dimensions.




\section{Summary and Discussion}

In this paper we have discussed the behavior of a spin system with
short range interactions in a random external field coupled
to the order parameter. If the distribution of the external field
is discrete, the resulting dynamics is chaotic due to the competition
between different equilibrium states of the system. We proposed
a general formalism for calculating the stationary and dynamical
properties of randomly driven systems and applied it to the Ising
model. In the mean-field approximation the stationary distribution
of the magnetization displays a spontaneous symmetry breaking
phase at low fields and temperatures. The transition between the
disordered and the ferromagnetic phase is first order 
and corresponds to a tangential bifurcation of the underlying map.
Close to the phase transition the characteristic relaxation time
diverges {\it exponentially} - leading to dynamic freezing. 
Depending on the control parameters, 
the stationary magnetization distribution can be a normal, multifractal or 
fat-fractal distribution in both
the disordered and ordered phases.



Our interest in this problem arises mainly in connection to 
understanding the nature of open systems with many degrees of freedom. 
Information processing systems, natural or artificial, have a macroscopic
number of connected elements
subject to external stimuli changing faster than the characteristic
thermal relaxation time. As illustrated by the simple example presented 
in this paper, such systems might
develop stationary states far from equilibrium which might be many times more
effective in dynamically storing information than simple thermal equilibrium
states. 
In this respect it would be also of interest to consider other choices
for the driving field distribution.
Continuous distributions, for instance, might lead
to very different stationary phase transitions than the one discussed
here.

In ``The randomly driven Ising ferromagnet'', Part II, we discuss the RDIM in
one and two dimensions.


\section*{Acknowledgements}
This article was initiated  
during PR's visit at the Hong Kong University of Science and Technology.
PR thanks the staff of the Department of Physics and in particular
N. Cue,  K.Y. Szeto, and M. Wong for their warm hospitality.
This work was partly supported by the DFG through SFB 517.





\end{document}